\documentclass[a4paper,aps,prd,showpacs,twocolumn,nofootinbib,10pt]{revtex4-1}
\pdfoutput=1
\usepackage{bbold}
\usepackage{mathtools}
\usepackage{ulem}

\usepackage{comment}
\usepackage{color}

\def\laq{~\raise 0.4ex\hbox{$<$}\kern -0.8em\lower 0.62ex\hbox{$\sim$}~}
\def\gaq{~\raise 0.4ex\hbox{$>$}\kern -0.7em\lower 0.62ex\hbox{$\sim$}~}

\def\beq{\begin{equation}}
\def\eeq{\end{equation}}
\def\bea{\begin{eqnarray}}
\def\eea{\end{eqnarray}}
\def\bean{\begin{eqnarray*}}
\def\eean{\end{eqnarray*}}

\def\lan {\langle}
\def\ran {\rangle}

\newcommand{\class}{{\sc class}}

\def \pa {\partial}
\def \ra {\rightarrow}

\def \la {\lambda}

\def \De {\Delta}
\def \de {\delta}

\def \al {\alpha}

\def \ka {\kappa}
\def \Ga {\Gamma}
\def \ga {\gamma}

\def \Sg {\Sigma}
\def \de {\delta}

\def \om {\omega}
\def \Om {\Omega}

\def \lap {\triangle}

\newcommand{\bel}{{\ell\kern -0.45em \ell}}
\newcommand{\Bell}{\boldsymbol{\ell}}

\def\laq{~\raise 0.4ex\hbox{$<$}\kern -0.8em\lower 0.62ex\hbox{$\sim$}~}
\def\gaq{~\raise 0.4ex\hbox{$>$}\kern -0.7em\lower 0.62ex\hbox{$\sim$}~}

\def\be{\begin{equation}}
\def\ee{\end{equation}}
\def \ga {\gamma}

\def\beq{\begin{equation}}
\def\eeq{\end{equation}}
\def\bea{\begin{eqnarray}}
\def\eea{\end{eqnarray}}

\def \pa {\partial}
\def \ra {\rightarrow}

\def \bx {\boldsymbol{x}}
\def \bn {\boldsymbol{n}}
\def \bnabla {\boldsymbol{\nabla}}

\newcommand{\Acal}{\mathcal A}
\newcommand{\Bcal}{\mathcal B}
\newcommand{\DD}{\mathcal D}
\newcommand{\Ecal}{\mathcal E}

\newcommand{\id}{\mathbb{1}}

\newcommand{\bfe}{{ \boldsymbol{e} }}

\newcommand{\bs}{{\boldsymbol{s}}}
\newcommand{\bv}{{\boldsymbol{v} }}
\newcommand{\by}{{\boldsymbol{y} }}

\newcommand{\dd}{\partial}

\def\laq{~\raise 0.4ex\hbox{$<$}\kern -0.8em\lower 0.62ex\hbox{$\sim$}~}
\def\gaq{~\raise 0.4ex\hbox{$>$}\kern -0.7em\lower 0.62ex\hbox{$\sim$}~}

\def\beq{\begin{equation}}
\def\eeq{\end{equation}}
\def\bea{\begin{eqnarray}}
\def\eea{\end{eqnarray}}
\def\bean{\begin{eqnarray*}}
\def\eean{\end{eqnarray*}}

\def \pa {\partial}
\def \ra {\rightarrow}

\def \la {\lambda}

\def \ka {\kappa}
\def \Ga {\Gamma}
\def \ga {\gamma}

\def \Sg {\Sigma}

\def \om {\omega}
\def \Om {\Omega}

\def \lap {\triangle}

\begin{document}

\title{Rotation of the CMB polarisation by foreground lensing}

\author{Enea~Di~Dio$^{a,b}$, Ruth~Durrer$^c$, Giuseppe~Fanizza$^d$ and Giovanni~Marozzi$^e$}

\affiliation{$^a$ Physics Division, Lawrence Berkeley National Laboratory, Cyclotron Rd, Berkeley, CA 94720
\\
$^b$ Berkeley Center for Cosmological Physics and Department of Physics, University of California, Berkeley, CA 94720
\\
$^c$ Universit\'e de Gen\`eve, D\'epartement de Physique Th\'eorique and CAP,
24 quai Ernest-Ansermet, CH-1211 Gen\`eve 4, Switzerland
\\
$^d$ Istituto Nazionale di Fisica Nucleare, Sezione di Pisa,
Largo B. Pontecorvo 3, 56127 Pisa, Italy
\\
$^e$ Dipartimento di Fisica, Universit\`a di Pisa, and
Istituto Nazionale di Fisica Nucleare,
Sezione di Pisa, Largo B. Pontecorvo 3, 56127 Pisa, Italy
}

\date{\today}

\begin{abstract} 
We investigate the weak lensing corrections to the CMB polarization anisotropies.
We concentrate on the effect of rotation and show that the rotation of polarisation is a true physical effect which has to be taken into account at second order in perturbation theory.  We clarify inconsistencies on the treatment of this rotation in the recent literature. We also show that at first order in perturbation theory there is no rotation of polarisation also for vector and tensor modes.

\end{abstract}


\maketitle

\section{Introduction}
\label{Sec1}
Cosmology has made enormous progress in the last twenty years from an order of magnitude into a precision science. At a closer look, much of this is due to the Cosmic Microwave Background (CMB). There is a wealth of high precision data on the anisotropies and the polarisation of the CMB, to a large extent dominated at present by the Planck data~\cite{Adam:2015rua,Planck:2015xua}. Moreover, there are also other important experiments which are more precise on smaller scales~\cite{Keisler:2015hfa,Array:2015xqh,Crites:2014prc,Naess:2014wtr}. CMB data are so precious because we understand them very well, see e.g.~\cite{RuthBook},  which comes from the fact that we can use, to a large extent, linear perturbation theory to interpret  them. Since the initial perturbation spectrum from inflation is simple, we can estimate the cosmological parameters,  on which the linear transfer functions sensitively depend, with high precision.

The main non-linearity which is relevant in our understanding of the CMB is lensing: due to the presence of massive foreground structures, CMB photons are deflected and arrive at the observer from a direction which does not agree with the direction of emission, see~\cite{Lewis:2006fu} for a review of CMB lensing. Not that if the CMB would be perfectly isotropic, $\de T\equiv 0$, lensing would have no effect of the CMB. In this sense CMB lensing goes beyond linear perturbation theory, we need both, temperature fluctuations and fluctuations in the foreground geometry. On small scales lensing is quite important and it changes the inferred fluctuation and polarisation spectra by 10\% and more at harmonics $\ell\gtrsim 1000$, corresponding to angular scales $\theta<\pi/1000 \simeq 10'$. This lensing of  CMB anisotropies and polarisation has been observed in several experiments, see e.g.~\cite{vanEngelen:2014zlh,Ade:2015zua,Ade:2015nch,Baxter:2016ziy}.

The importance of the effect has prompted several of us to investigate whether { higher} order contributions to it might be relevant for future, high precision S4 type~\cite{Abazajian:2016yjj} experiments or future satellites~\cite{Delabrouille:2017rct,DiValentino:2016foa,Suzuki:2018cuy}. In the standard treatment the contributions from the first order deflection angle are 'summed up' assuming Gaussianity. Including this non-linearity, which is standard in present CMB codes like CAMB~\cite{Lewis:1999bs} or {\sc class}~\cite{Lesgourgues:2011re,Blas:2011rf}, is relevant for a precise analysis of  recent experiments like Planck.
However, in this treatment, the deflection angle is always calculated in the so-called Born approximation, i.e.~by integrating the lensing potential along the unperturbed photon geodesic.
At second order this is no longer correct and a treatment beyond the Born approximation is in principle requested. Recent works~\cite{Pratten:2016dsm,Marozzi:2016uob,Lewis:2016tuj,Marozzi:2016qxl,Marozzi:2016und,Bohm:2016gzt,Lewis:2017ans,Fabbian:2017wfp,Takahashi:2017hjr,Bohm:2018omn,Beck:2018wud}  have considered this and  other effects including, in several steps, most  higher order contributions to CMB lensing.

In most of their calculations, presented in Refs.~\cite{Pratten:2016dsm,Marozzi:2016uob,Lewis:2016tuj,Marozzi:2016qxl,Marozzi:2016und,Lewis:2017ans}, the results of the two groups involved {in the analytic and numerical evaluation of the higher-order effects} are in reasonable agreement, but there is one exception which is the subject of the present work: in principle, parallel transport can lead to a rotation of the Sachs basis, i.e. the orthonormal basis on the 'screen' normal to the photon direction and to the four velocity of the observer, by an angle { which we can call
 $\alpha$}. 
In this case the polarisation tensor rotates by { $-\alpha$ and changes the complex polarisation $P=Q+i U$ by $P\mapsto \exp(-2i\alpha)P$} which affects the polarisation spectrum and especially induces $B$-polarisation from an original $E$-polarisation spectrum. 
$B$-polarisation is already induced by the effect of re-mapping by lensing at first order and this has been measured by several experiments~\cite{Ade:2014afa,Array:2015xqh,Ade:2015nch}. 
This rotation could reduce the de-lensing efficiency of gradient based methods~\cite{Hirata:2003ka},
and therefore the sensitivity of next generation of CMB experiments to the tensor-to-scalar ratio.

In particular, using the so-called geodesic light-cone (GLC) gauge~\cite{Gasperini:2011us,Fanizza:2013doa,Fanizza:2015swa}, in Ref.~\cite{Marozzi:2016qxl} it has been estimated that the contribution of $B$-modes from rotation is at the percent level for $\ell>2500$. This effect induced by the rotation field could, therefore, affect in a non-negligible way the reconstruction of the tensor-to-scalar ratio $r$ for future ex\-peri\-ments~\cite{Abazajian:2016yjj,Delabrouille:2017rct}. If this is correct, this rotation is of uttermost importance for the analysis of these experiments. However, in Ref.~\cite{Lewis:2017ans} the authors  show that, in longitudinal gauge, the Sachs basis does not rotate. In fact, we shall show that in this gauge, a spatial vector normal to the photon direction does not rotate at any order when parallel transported along the photon geodesic in a quasi-Newtonian gravitational potential.
The authors of Ref.~\cite{Lewis:2017ans} therefore argue that higher order lensing effects on the CMB polarisation are very small and can be safely neglected in the analysis of planned experiments. 
At first sight, this suggests that the rotation angle of the polarization tensor parallel transported from the last surface scattering to the observer may be gauge-dependent.

This is the present state of affairs. The CMB power spectra, however, are observables and cannot depend on the coordinate system which is used to compute them. Therefore, either the conclusion of Ref.~\cite{Lewis:2017ans} or the one of Ref.~\cite{Marozzi:2016qxl} (or both) must be wrong. The important question is: rotation with respect to what is relevant for the CMB spectra?  As already discussed in \cite{Marozzi:2016qxl} (see beginning of Section VI), this cannot be the rotation with respect to some arbitrarily chosen coordinate system, but it must be a physically defined rotation. 

In the next section we show that the relevant rotation is the one of the  Sachs basis with respect to the direction of a vector connecting neighbouring geodesics. Since this geodesic deviation vector is Lie transported along the photon geodesic, this means that the relevant rotation angle $\alpha$ is the change in the angle between a Lie transported and a parallel transported vector in the screen.
In more detail, as we shall  clearly show in the next section, in an arbitrary coordinate system the physical angle  $\al$
is given by the sum $\beta+\omega$, 
where $\beta$ is the rotation of the Sachs basis with respect to an arbitrary fixed basis while $\om$ gives the rotation of the geodesic deviation vector in the fixed basis (hence the change of the angle between the parallel transported polarisation direction and the direction of the geodesic deviation vector is $-\beta-\om=-\al$).
We also show that,  for scalar perturbation,  this is exactly the angle $\beta$ calculated in~\cite{Marozzi:2016qxl} to second order\footnote{According to the definition made here in \cite{Marozzi:2016qxl} $\alpha$ has been identified with $\beta$ (see Sect. \ref{Sec2}).}.
In longitudinal gauge, the parallel transported Sachs basis does not rotate but the geodesic deviation vector of neighbouring photon geodesics rotates at second order by the angle $\om$ given by the amplification matrix,  ($\partial {\bf n}/\partial {\bf n'}$) (where ${\bf n}$ is the incoming photon direction and 
${\bf n'}$ is the source direction) of the lens map at second order. Denoting this angle in longitudinal gauge by $\om_{LG}$, this is consistent with the finding 
$\alpha_{GLC} =\om_{LG}$ of~\cite{Marozzi:2016qxl}.

The remainder of this work is structured as follows: in Section~\ref{Sec2} we derive the relation between the angles $\om$, $\beta$ and the CMB polarisation power spectra. In Section~\ref{Sec3} we show that in first order perturbation theory  $\beta+\om$ vanishes not only for scalar but also for vector and tensor perturbations. In Section~\ref{Sec4} we calculate this rotation angle to second order for scalar perturbation in longitudinal gauge where we find again the result derived in Ref.~\cite{Marozzi:2016qxl}.
In Section~\ref{Sec5} we briefly discuss our findings and conclude. 
In Appendix~\ref{app:A} we give some technical details, while 
 in Appendix~\ref{app:B} we show the equivalence between the derivation presented in this manuscript and the previous one in Ref.~\cite{Marozzi:2016qxl}. In Appendix~\ref{app:C} we present analytic approximations for the slope of the spectra at high and low $\ell$.
 
All the calculations are performed in the flat sky approximation, see~\cite{Lewis:2006fu,RuthBook}, which is  
 sufficient if we are interested in spherical harmonics with $\ell>50$. 
 Furthermore, numerical results are obtained within the Limber approximation \cite{Limber:1954zz,LoVerde:2008re}, 
which works very well for CMB lensing (CMB lensing is appreciable only for $\ell \ge 100$, where the Limber approximation is very close to the exact solution).
We also do not consider effects from lensing beyond the Born approximation which are not related to lensing-rotation. This mainly because on these effects the references~\cite{Pratten:2016dsm,Lewis:2016tuj,Lewis:2017ans} and \cite{Marozzi:2016uob,Marozzi:2016qxl,Marozzi:2016und} agree reasonably well. But also since for high $\ell$ rotation is the dominant effect on the $B$-polarisation spectrum. We also do not discuss here the dominant lensing terms which can be obtained within the Born approximation as these are well known, see e.g.~\cite{Lewis:2006fu,RuthBook} and there is no controversy concerning these terms.
 
\section{CMB spectra}
\label{Sec2}
Throughout we shall work in the flat sky approximation which is fully sufficient for harmonic modes $\ell> 100$.
Let us consider two points in the sky, $\bx$ and $\bx'$. They may have a slightly different temperature and different polarisation. Here we are interested in the polarisation. Since Thomson scattering only produces linear polarisation, we expect the Stokes parameter $V$ to vanish and introduce the complex polarisation 
\be
P(\bx) = Q(\bx)+iU(\bx) 
\ee
which of course depends on the orientation of our coordinate system. $P$ has helicity 2 and transforms under a rotation of the basis by an angle $\theta$ as $P(\bx) \mapsto \exp(-2i\theta)P(\bx)$.
In Fourier space we can express $P$, and its complex conjugate $P^*=Q-iU$, in terms of $E$ and $B$ polarisations as~\cite{Seljak:1996ti}
\bea \label{e:P-EB}
P(\bx) &=&Q(\bx) +iU(\bx) 
\nonumber \\
 &=& -\int\frac{d^2\Bell}{2\pi}\left[E(\Bell)+iB(\Bell)\right]e^{2i\varphi_\ell}e^{i\Bell\cdot\bx}\, , \\
P^*(\bx) &=&Q(\bx) -iU(\bx) 
\nonumber \\
 &=& -\int\frac{d^2\Bell}{2\pi}\left[E(\Bell)-iB(\Bell)\right]e^{-2i\varphi_\ell}e^{i\Bell\cdot\bx}\,,   \label{e:Pbar-EB}
\eea
where $\varphi_\ell$ is the polar angle of the 2d vector $\bel$.
Inversely
\bea\label{e:E+iB}
E(\bel)+iB(\bel) &=& -\int\frac{d^2\bx}{2\pi}P(\bx) e^{-2i\varphi_x}e^{-i\Bell\cdot\bx}\, ,\\
E(\bel)-iB(\bel) &=& -\int\frac{d^2\bx}{2\pi}P ^*(\bx)e^{2i\varphi_x}e^{-i\Bell\cdot\bx}\,, \label{e:E-iB}
\eea
where $\varphi_x$ is the polar angle of  $\bx$ (see, e.g.~\cite{RuthBook}). 
In the above integrations one can of course fix the non-integrated variable along the abscissa axis so that $\varphi$ denotes the angle between $\bel$ and $\bx$ in both cases.

In the flat sky approximation the power spectra of the $E$- and $B$-polarisation are defined by
\bea\label{e:ClE}
\langle E(\Bell)E^*(\Bell')\ran &=& \de(\Bell-\Bell')C_\ell^{E} \,, \\
\lan B(\Bell)B^*(\Bell')\ran &=& \de(\Bell-\Bell')C_\ell^{B} \,. \label{e:ClB}
\eea
The Dirac delta function is a consequence of statistical isotropy (which correponds to statistical homogeneity on the flat sky) and we request statistical parity invariance so that the correlations between $E$ and $B$ vanish.

We now want to correlate $P(\bx)$ with $P(\bx')$ which, by statistical isotropy depends only on $\bs =\bx-\bx'$. In order to define a correlation function which is independent of the orientation of the basis $({\bf e}_1,{\bf e}_2)$, we determine the polarisation with respect to  a new basis $({\bf e}'_1,{\bf e}'_2)$ with ${\bf e}'_1=\hat \bs$, the unit vector in direction $\bs$. This new polarisation is then given by
\be\label{e:rotP}
P_s(\bx)=e^{-2i\varphi_s}P(\bx)\,,
\ee
where $\varphi_s$ is the polar angle of $\bs$ with respect to the original basis $({\bf e}_1,{\bf e}_2)$.
With respect to the new intrinsic basis $({\bf e}'_1,{\bf e}'_2)$ we now define
\bea\label{e:xi-rot+}
\xi_+(s) &=& \lan P^*_s(\bx)P_s(\bx')\ran = \lan P^*(\bx)P(\bx')\ran = 
\nonumber \\
&=&
 \lan Q(\bx)Q(\bx')\ran +  \lan U(\bx)U(\bx')\ran \,, \\
\xi_-(s) &=& \lan P_s(\bx)P_s(\bx')\ran =   \lan e^{-4i\varphi_s}P(\bx)P(\bx')\ran \nonumber\\
     &=& \lan Q_s(\bx)Q_s(\bx')\ran-  \lan U_s(\bx)U_s(\bx')\ran \,.  \label{e:xi-rot}
\eea
The terms $\lan Q_s(\bx)U_s(\bx')\ran$ vanish since they change sign under parity, $\bs\ra-\bs$ and we assume statistical  parity invariance of the signal.
In terms of the $E$ and $B$ power spectra we obtain
\bea\label{e:xi+EB}
\xi_+(s) &=& \frac{1}{2\pi}\int_0^\infty d\ell\ell\left[C_\ell^{E}+ C_\ell^{B}\right]J_0(\ell s) \,, \\
\xi_-(s) &=& \frac{1}{2\pi}\int_0^\infty d\ell\ell\left[C_\ell^{E}- C_\ell^{B}\right]J_4(\ell s)\,. 
\label{e:xi-EB}
\eea 
For this we simply use Eqs.~(\ref{e:P-EB}, \ref{e:Pbar-EB}) and (\ref{e:ClE}, \ref{e:ClB}), as well as the identity
$$ \frac{(-i)^n}{2\pi}\int_0^{2\pi}d\varphi \,\, e^{ni\varphi + iy\cos\varphi} = J_n(y) \,,$$
 where $J_n$ denotes the Bessel function~\cite{Abram} of order $n$.
Eqs.~(\ref{e:xi+EB}, \ref{e:xi-EB}) are  readily inverted to 
\bea\label{e:CE+CB}
C_\ell^{E}+ C_\ell^{B} &=& {2\pi}\int_0^\infty ds\,s\,\xi_+(s)J_0(\ell s) \,,\\
C_\ell^{E}- C_\ell^{B}&=&{2\pi}\int_0^\infty ds\,s\,\xi_-(s) J_4(\ell s) \,. \label{e:CE-CB}
\eea
 \begin{figure}[t!]
\centering
\includegraphics[width=0.9\linewidth]{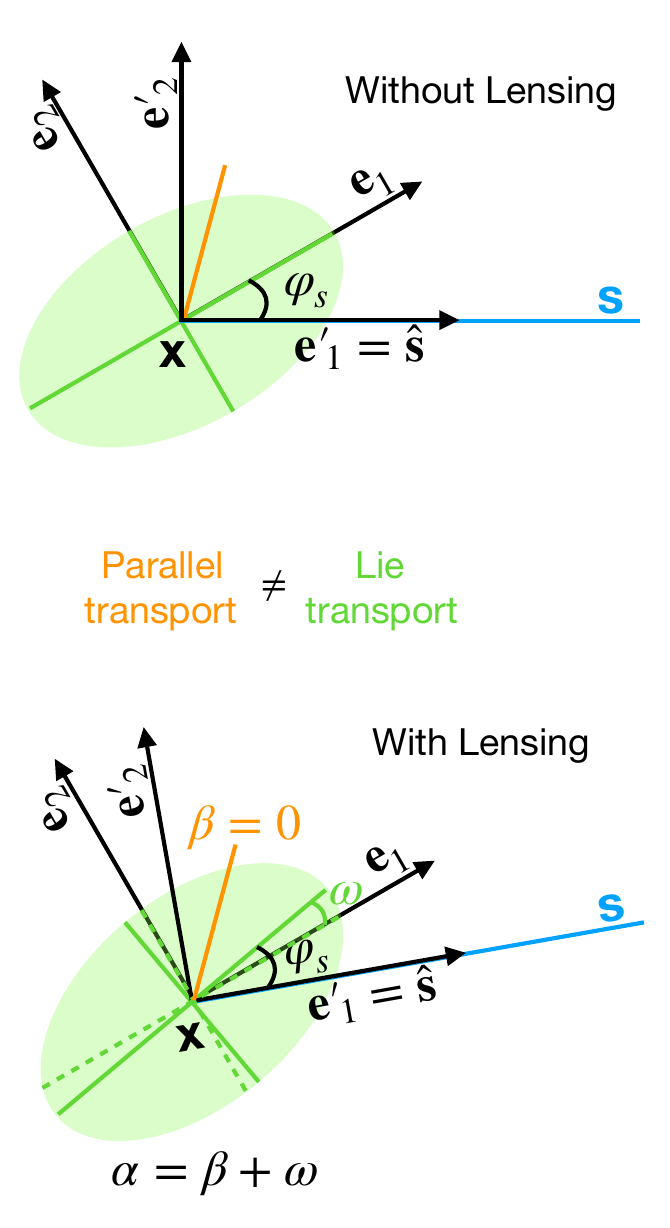}
\caption{We show schematically the rotation of the polarization for scalar perturbations in Poisson gauge. The polarization vector (orange) is parallel transported from the source to the observer plane and does not rotate wrt the arbitrary basis $\left\{ {\bf e}_1 , {\bf e}_2 \right\}$, here chosen to align with an unlensed elliptical image ($\beta = 0$). The image (green) and the separation vector $\bs$ (with the basis $\left\{ {\bf e}'_1 , {\bf e}'_2 \right\}$) are Lie transported and they rotate by an angle $\omega$. The physical effect is due to the gauge invariant angle $\alpha = \beta + \omega$ which is generated by the different rotations induced by a parallel transport with respect to a Lie transport.} 
\label{fig:picture}
\end{figure}
We first concentrate on the rotation of the polarisation only. We shall see that this contribution dominates.
Let us  assume that at position $\bx$ lensing rotates the polarisation basis by parallel transport by some angle $\beta(\bx)$, $({\bf e}_1+i{\bf e}_2) \mapsto \exp(i\beta(\bx))({\bf e}_1+i{\bf e}_2)$.  Furthermore, vectors or tensors have to be transformed with the Jacobian of the lens map\footnote{By this we mean the standard fact that under a transformation $x\mapsto\phi(x)$, vectors transform with $\bv(x)\mapsto \phi^*(x)\bv(\phi(x))$, where $\phi^*$ denotes the tangent map.}. Here we only consider the rotation, $\om$, by which  the unlensed tangent vectors have to be transformed. Hence  the system $({\bf e}'_1+i{\bf e}'_2)$ describing the basis along the unlensed direction $\bs$  at $\bx$ is rotated by $\om(\bx)$ $({\bf e}'_1+i{\bf e}'_2) \mapsto \exp(i\om(\bx))({\bf e}'_1+i{\bf e}'_2)$.  Introducing the  relative angle $\al=\beta+\om$,
we infer from Eq.~(\ref{e:rotP})  that the polarisation oriented with respect to $({\bf e}'_1,{\bf e}'_2)$  is rotated by $-\al$ (and correspondingly for $\bx'$). 
The different angles are represented in Fig.~\ref{fig:picture} for the case of scalar perturbations in Poisson gauge (see Sections \ref{Sec3} and \ref{Sec4}).
Denoting the rotated polarisation by $\tilde P^r$ we have
\bea
\tilde P^r_s(\bx) &=& e^{-2i\al(\bx)}e^{-2i\varphi_s}P(\bx) \,\\
\tilde P^r_s(\bx') &=& e^{-2i\al(\bx')}e^{-2i\varphi_s}P(\bx') \,.\\  \nonumber
\eea

Correspondingly, the lensed correlation functions $\xi_{\pm}$ pick up a factor $\exp(-2i(\al(\bx)\mp\al(\bx')))$ from a rotation. It is this and only this angle $\al=\om+\beta$ which is truly observable. The angle by which a tangent vector is rotated by the lens map, i.e. the rotation of a vector connecting two neighbouring geodesics, which is Lie transported, and the Sachs basis which is parallel transported.
Note that for clarity we take into account here only the rotation induced by lensing, neglecting all other effects from lensing even though they may be significantly larger.

More precisely
\bea
\tilde \xi^r_+(s)&=&\lan e^{+2i(\al(\bx)-\al(\bx'))}\ran\xi_+(s) \,, \\
\tilde \xi^r_-(s)&=&\lan e^{-2i(\al(\bx)+\al(\bx'))}\ran\xi_-(s) \,.
\eea
Here we assume that we can neglect the correlations of the unlensed polarisation (mainly produced at the last scattering surface) and the deflection angles (generated by foreground structures). We now use a relation which is strictly true only for Gaussian variables, but we assume that the non-Gaussianity of $\al$ gives a subdominant contribution with respect to the leading order effect. 
Assuming Gaussianity we can set
\bea
\lan e^{+2i(\al(\bx)-\al(\bx'))}\ran & \simeq& e^{-2 \lan(\al(\bx)-\al(\bx'))^2\ran}
\nonumber \\
 &=&e^{-4(C_\al(0)-C_\al(s))}\, , \
 \label{exp_xi_plus} \\
\lan e^{-2i(\al(\bx)+\al(\bx'))}\ran &\simeq & e^{-2 \lan(\al(\bx)+\al(\bx'))^2\ran} 
\nonumber \\
&=&e^{-4(C_\al(0)+C_\al(s))}\, ,  
\label{exp_xi_minus}
\eea
where we have introduced the correlation function of the angle $\al$,
$$
C_\al(s) =\lan\al(\bx)\al(\bx')\ran \,, \qquad s=|\bx-\bx'| \,.
$$
Using Eqs.~(\ref{e:CE+CB}, \ref{e:CE-CB}), we then find  the following result for rotation induced $B$-spectrum
\begin{widetext}
\bea
\tilde C^{Br}_\ell &=& \pi\int_0^\infty ds\, s[\tilde\xi^r_+(s)J_0(\ell s) -\tilde\xi^r_-(s)J_4(\ell s)]
=\pi\,e^{-4C_\al(0)}\int_0^\infty ds\, s[\xi_+(s)J_0(\ell s)e^{4C_\al(s)} -\xi_-(s)J_4(\ell s)e^{-4C_\al(s)}]\nonumber\\
&=&\frac{1}{2}\,e^{-4C_\al(0)}\int_0^\infty ds\, s\int_0^\infty d\ell'\ell'C_{\ell'}^{E}[J_0(\ell' s)J_0(\ell s)e^{4C_\al(s)} -J_4(\ell' s)J_4(\ell s)e^{-4C_\al(s)}]\nonumber\\
&\approx&\frac{1}{2}\int_0^\infty ds\, s\int_0^\infty d\ell'\ell'C_{\ell'}^{E}[J_0(\ell' s)J_0(\ell s)e^{4C_\al(s)} -J_4(\ell' s)J_4(\ell s)e^{-4C_\al(s)}]
\label{eq:lensB}
\eea
\end{widetext}
where we have assumed no primordial $B$-modes. The second line of Eq.~\eqref{eq:lensB} shows that the correction due to the variance of $\alpha$, $C_\al(0)$ factorizes  and contributes to a constant shift in the spectrum of the $B$-modes. This  confirms what has been found in \cite{Marozzi:2016qxl} at the leading order, where this variance has been estimated to be of order $10^{-4}-10^{-5}$ depending on the spectrum used. This justifies the approximation made in the last line of Eq.~\eqref{eq:lensB}. For a practical calculations, the exponentials in the integrand of Eq.~\eqref{eq:lensB} can be further expanded, giving the leading result
\be \label{rotated_Bmode}
\tilde C^{Br}_\ell \!\!
  = \!\!2\!\! \int_0^\infty \!\!\!\! \!\!ds sC_\al(s) \!\! \int_0^\infty \!\!\!\!\!\!d\ell'\ell'C_{\ell'}^E[J_0(\ell s) J_0(\ell' \!s)+J_4(\ell s) J_4(\ell' \!s)] \,,
\ee
where we have used
\be
\int_0^\infty ds s J_n(\ell s) J_n(\ell' s) = \frac{1}{\ell}\de(\ell-\ell') \,.
\ee
With similar manipulations we find the rotation of the $E$-spectrum,
\bea
&&\tilde C^{Er}_\ell= [1-4C_\al(0)]C^E_\ell 
\nonumber \\
&&+2\! \! \int_0^\infty \!\!\!  \! ds sC_\al(s) \!\!\int_0^\infty \!\!\!\!d\ell'\ell'C_{\ell'}^E[J_0(\ell s) J_0(\ell' s)-J_4(\ell s) J_4(\ell' s)] \,.
\nonumber \\
\eea

From our derivation it becomes  clear that the polarisation $P_s$ which enters the calculation of the $E$ and $B$ power spectra depends on $\al$ and is rotated by it when $\bs$ or/and ${\bf e}_1$ is rotated. Whether one or the other or both are rotated depends on the chosen coordinate system, but the sum of the two rotation angles, the total rotation, is physical. In previous calculations of one group~\cite{Pratten:2016dsm,Lewis:2016tuj,Lewis:2017ans} longitudinal gauge (LG) was used  where, as we shall see, $\beta$ vanishes but $\om$ is non-zero at second order. In the calculations of the other group~\cite{Marozzi:2016uob,Marozzi:2016qxl,Marozzi:2016und}
geodesic light-cone (GLC) gauge \cite{Gasperini:2011us} was used, where directly the angle $\al$ of the rotation of the Sachs basis,  $({\bf e}_1, {\bf e}_2)$, with respect to the incoming photon direction is determined. There it has been shown that  $\al_{\rm GLC}=\om_{\rm LG}$ at second order  for scalar perturbations.

Indeed, according to what we have shown so far, the rotation between the displacement vector of two null geodesics $\xi^\mu$ and the polarisation vector $\epsilon^\mu$ of a photon is given by
\beq
\cos\alpha=\hat\xi^\mu\hat\epsilon_\mu=\hat\xi^A\hat\epsilon_A
\label{eq:cornerstone}
\eeq
where hatted quantities are normalized vectors and $\hat\xi^A\equiv\hat\xi^\mu e_\mu^A$ and $\hat\epsilon^A\equiv\hat\epsilon^\mu e_\mu^A$ and $e^A_\mu$ is the Sachs basis. Because both $\hat \epsilon^\mu$ and $e^A_\mu$ are parallel transported, we have that $\hat\epsilon^A$ remains constant while traveling along the geodesic. This means that any change in $\alpha$ can only be due to the rotation of $\hat\xi^A$. Because photons on the same light-cone travel at fixed angular coordinates\footnote{The constancy of these angular coordinates leads to equal them with the photon's incoming direction as seen by the observer. This explicit identification has been recently discussed and implemented in \cite{Mitsou:2017ynv,Fanizza:2018tzp} by exploiting the residual gauge freedom of the GLC coordinates.} and start at the same time, their displacement vector will be $\xi^\mu= \delta^\mu_a C^a$, with $C^a$ constant and $a=1,2$ runs over the angular coordinates which do not change in GLC gauge (see \cite{Fanizza:2013doa} for a detailed derivation). Because of this, in GLC, $\xi^A=e^A_a C^a$. From Eq. \eqref{eq:cornerstone} and due to the constancy of $\hat\epsilon$, it follows that in GLC $\alpha$ changes according to the rotation of the Sachs basis. This angle is what has been found in Appendix C of \cite{Marozzi:2016qxl}.
\\
Let us underline that $\alpha$ is a gauge invariant quantity such that its value is independent of the coordinate system. The equality between its value and the rotation of the basis is a peculiarity of the GLC gauge where $\xi^\mu$ is somehow trivial.

It is convenient to rewrite Eq.~\eqref{rotated_Bmode} by introducing the angular power spectrum of the rotation angle $\alpha$ as
\be \label{beta_spectrum}
C^{\alpha \alpha}_\ell = 2 \pi \int ds   s C_\alpha \left( s \right) J_0\left( s \ell \right).
\ee
Simply using the inverse relation
\be
C_\alpha \left( s \right) = \frac{1}{2 \pi} \int d\ell  \ell C_{\ell}^{\alpha\alpha} J_0\left( s \ell \right) \,,
\ee
we find
\be \label{rot_B_eq}
\tilde C^{Br}_\ell =\frac{1}{\pi} \int d\ell' d\ell'' \ell' \ell'' C_{\ell''}^{\alpha\alpha} C_{\ell'}^E F_{\ell \ell' \ell''} \,,
\ee
where we have introduced the geometrical factor (see Appendix~\ref{app:A} for further details)
\be \label{F_factor}
F_{\ell \ell' \ell''} = \int ds s J_0 \left( \ell'' s \right) \left[J_0(\ell s) J_0(\ell' s)+J_4(\ell s) J_4(\ell' s)\right] \, .
\ee

Before we go on, we also want to find the effect from the rotation of the position $\bx$. For this we write the relation between the unrotated $\tilde \bx$ and the rotation $\bx$ positions\footnote{ For the purpose of this derivation we neglect the gradient part in the deflection angle. Due to the different parity the gradient and the curl components of the deflection field are uncorrelated.} as 
\be
\tilde \bx = \bx + \bnabla \wedge \Omega \,.
\ee  
Here $\Om$ is the  potential of the rotation angle and~$\al=-\De\Om/2$. Note that in 2 dimensions 
$$  (\bnabla \wedge \Omega)^a = \epsilon^{ab}\nabla_b\Om\,,$$
where $\epsilon^{ab}$ is the totally antisymmetric symbol in 2 dimensions and $\Om$ is a (pseudo-)scalar. 
In the literature this has been considered mainly in longitudinal gauge where for scalar perturbations $\al=\om$. But as we have argued before, this expression is not gauge invariant and we have to consider the rotation of $
\bx$ with respect to the Sachs basis which is given by the angle $\al$. As we shall show in the next section, this implies that the curl component of the deflection field can not be sourced by any linear perturbation, including vector and tensor perturbations contrarily to what claimed in the past literature, starting from Refs.~\cite{Stebbins:1996wx,Dodelson:2003bv}.
We denote the so displaced polarisation by $\tilde P^d(\bx)$.
To lowest order in $\al$ and hence $\Om$ this changes the polarization into
\be
\tilde P^d(\bx) = P(\tilde \bx)  = P(\bx)+  \epsilon^{ab}\nabla_b\Om (\bx)\nabla_aP(\bx) \,.
\ee
Inserting this in the expressions \eqref{e:xi-rot+} and \eqref{e:xi-rot}  for $\xi_+(s)$ and $\xi_-(s)$,  and Fourier transforming\footnote{We remark that in our Fourier convention, the angular power spectrum is the Fourier transform of the correlation function up to a factor $2\pi$, i.e.~$\xi(\ell) = C_\ell/ (2\pi)$.}, we find
\bea
\tilde \xi_+^d(\ell) &=&\int \frac{d^2\ell'}{(2\pi)^3}(\bel\wedge\bel')^2C_{|\Bell-\Bell'|}^\Om C_{\ell'}^E \, , \\
\tilde \xi_-^d(\ell) &=& \int \frac{d^2\ell'}{(2\pi)^3}(\Bell\wedge\Bell')^2C_{|\Bell-\Bell'|}^\Om C_{\ell'}^Ee^{4i\varphi_{\Bell,\Bell'}} \, , \qquad
\eea
where $\varphi_{\Bell,\Bell'}= \varphi_{\Bell'}-\varphi_{\Bell}$.
Here we use the notation $\bx\wedge\by =\epsilon_{ab}x^ay^b$ for the vector product in 2 dimensions. Note that in two dimensions a vector product is a (pseudo-)scalar (the length of the corresponding radial vector in 3d).
Parity invariance $\xi(\bs)=\xi(-\bs)$ for all correlation functions implies that power spectra are real. The imaginary part above therefore cannot contribute and we may replace the exponential by its real part,
\begin{center}
$\text{Re}\left(e^{4i\varphi_{\Bell,\Bell'}}\right)= \cos(4\varphi_{\Bell,\Bell'})= 1-2\sin^2(2\varphi_{\Bell,\Bell'}) \,.
$
\end{center}
With this we find
\bea
\tilde C^{Bd}_\ell &=&\frac{2\pi}{2}\left( \tilde \xi_+^d(\ell) -\tilde \xi_-^d(\bel) \right) \nonumber \\
&=&
\int \frac{d^2\ell'}{(2\pi)^2}(\Bell\wedge\Bell')^2C_{|\Bell-\Bell'|}^\Om C_{\ell'}^E\sin^2(2\varphi_{\Bell,\Bell'}) \,. \qquad \label{e:Bcurl}
\eea
This term agrees exactly with the corresponding term in \cite{Hirata:2003ka}, (see Eq.~(9)) or \cite{Lewis:2017ans} (see Eq.~(B9)).

Interestingly also the cross term between the rotation of polarisation and the curl deflection does not vanish. Taking the first non-vanishing order in both rotation of polarisation and the curl deflection of position we find
\be\label{e:Pexp1}
\tilde P^{rd}(x) = P(x)-2i\al(x)P(x) + \epsilon^{ab}\nabla_b\Om(x)\nabla_aP(x) \,.
\ee
This leads to the following cross terms in $\tilde \xi_+(\ell)$ and  $\tilde \xi_-(\ell)$,
\bea
\tilde \xi^{rd}_+(\ell) &=&  0 \,, \\
\tilde \xi^{rd}_-(\ell) &=& -4\,i\int \frac{d^2\ell'}{(2\pi)^2} (\Bell\wedge\Bell')e^{4i\varphi_{\Bell,\Bell'}}C_{|\Bell-\Bell'|}^{\al\Om} C_{\ell'}^E \nonumber \\
&=& 4\!\int\! \frac{d^2\ell'}{(2\pi)^2} (\Bell\wedge\Bell')\sin(4\varphi_{\Bell,\Bell'})C_{|\Bell-\Bell'|}^{\al\Om} C_{\ell'}^E\,. \qquad  \label{e:xi-mixed}
\eea
In \eqref{e:xi-mixed} we have again only considered the non-vanishing real part.  The $B$-spectrum  hence acquires the cross term,
\be\label{e:mixed}
\tilde C^{Brd}_\ell =-2\!\int\! \frac{d^2\ell'}{(2\pi)^2}(\Bell\wedge\Bell')C^{\al\Om}_{|\Bell-\Bell'|} C_{\ell'}^E\sin(4\varphi_{\Bell,\Bell'}) \,. 
\ee
Also this expression would agree with the one in  \cite{Lewis:2017ans} (Eq.~(B9)) if their rotation angle $\beta$ would agree with our $\al$ which is not the case. Their $\beta$ is much smaller than our $\al$ and actually is due to an effect which we neglect in our treatment.

If we do not perform the integration over angles in \eqref{rotated_Bmode} and replace the correlation function $C_\al(s)$ with the corresponding power spectrum, or if we simply compute the $\al$-$\al$ contribution to the $B$-polarisation spectrum starting from (\ref{e:Pexp1}), we can write the effect from rotation to lowest order in a similar way. Following the same steps as for $\tilde C^{Bd}$ we find
\be\label{e:Brotpert}
\tilde C^{Br}_\ell =4\int \frac{d^2\ell'}{(2\pi)^2}C_{|\Bell-\Bell'|}^\al C_{\ell'}^E\cos^2(2\varphi_{\Bell,\Bell'}) \,. \qquad
\ee
This result agrees also with \eqref{eq.b2} (up to a variable transform $\Bell'\mapsto \Bell-\Bell'$).
Adding all the terms together  and using $\De\Om=-\al/2$ hence $C_{\ell}^\Om=4\ell^{-4}C_{\ell}^\al$ and $C_{\ell}^{\al\Om}=2\ell^{-2}C_{\ell}^\al$, we can write the total $B-$spectrum induced by rotation, to lowest order in the rotation angle $\al$, as
\begin{widetext}
\be\label{e:Btot}
\De\tilde C^{B}_\ell = 4\int\frac{d^2\ell'}{(2\pi)^2}C_{|\Bell-\Bell'|}^\al C_{\ell'}^E
\left[\cos^2(2\varphi_{\Bell,\Bell'})-\frac{\ell\ell'}{|\Bell-\Bell'|^2}\sin(4\varphi_{\Bell,\Bell'}) \sin(\varphi_{\Bell,\Bell'}) +\frac{(\ell\ell')^2}{|\Bell-\Bell'|^4}\sin^2(2\varphi_{\Bell,\Bell'})\sin^2(\varphi_{\Bell,\Bell'}) \right] \,.
\ee
\end{widetext}

In section \ref{Sec5} when we present numerical results for the contribution to the $B$-spectrum from rotation for second order scalar perturbations, we shall see that the term from the rotation of the polarisation dominates the total result on all scales.

\section{Lie transport and parallel transport at first order}\label{Sec3}
In this section we calculate to first order the relevant angle $\al=\omega+\beta$.
To determine $\om$ we study the propagation of neighbouring photons in an infinitesimal light beam (sometimes called a null congruence) which is given by the so called Jacobi matrix~$\DD$, see e.g.~\cite{Schneider:1992,Perlick:2004tq,Straumann:2013spu}. 
Denoting the geodesic deviation vector by $X$ we find from the geodesic deviation equation
\be\label{e:geodev}
\ddot X^\al =  -R^\al_{\beta\mu\nu}k^\beta X^\mu k^\nu \,.
\ee
Note that the geodesic deviation equation together with the geodesic equation for $k$ implies that $X$ is Lie transported\footnote{This is easily seen in coordinate free notation where $\ddot X=\nabla_k\nabla_kX$ and $-R^\al_{\beta\mu\nu}k^\beta X^\mu k^\nu =(-R(X,k)k)^\al = (\nabla_k\nabla_X-\nabla_X\nabla_k)k=\nabla_k\nabla_Xk$, where we have used $\nabla_kk=0$. Hence (\ref{e:geodev}) implies $\nabla_k(\nabla_kX-\nabla_Xk)= \nabla_k([k,X])=\nabla_kL_kX=0$, where $L_k$ denotes the Lie derivative in direction $k$. But in the source plane, $\la=\la_{\rm{in}}$, we can choose $X$ to denote a coordinate direction, $X=\dd_s$ and $k=\dd_\la$ so that at $\la_{\rm{in}}$ we have $[k,X]=[\dd_\la,\dd_s]=0$, so that constancy of $L_kX$ implies $L_kX=0$ along the photon geodesic.}.
Decomposing $X$ into a part parallel to $k$, a part parallel to the observer/emitter 4-velocity $u$ and
a part in the 'screen' normal to $k$ and $u$ with basis $({\bf e}_1,{\bf e}_2)$, we denote by $\DD$ the map which maps directions at the observer given by the screen  basis $({\bf e}_1,{\bf e}_2)$  to a distance vector $Y$ on the screen of the emitter. Since (\ref{e:geodev}) is linear, $\DD$ is a linear map which expresses the vector $Y$ in terms of the Sachs basis at the emitter, i.e.~the basis  $({\bf e}_1,{\bf e}_2)$ which is parallel transported backwards from the final to the initial screen,
$$ Y_a = {\DD_a}^b{e}_b + y_ak + z_au\,, $$
where $(a,b)$ take the values 1 and 2 and $Y_a$ starts out as $0$.
The  $2\times2$ matrix $\DD$ can be written as
\be\label{e:Jac}
\DD = R(\omega)R(\chi)\left(\begin{array}{cc} D_+ & 0 \\ 0 & D_-\end{array}\right)R(-\chi)
\ee
$$ \mbox{where }\qquad
R(\ga) = \left(\begin{array}{cc} \cos\ga & -\sin\ga \\ \sin\ga & \cos\ga\end{array}\right)
$$
denotes a rotation by an angle $\ga$. The matrix
$$
\Sg = R(\chi)\left(\begin{array}{cc} D_+ & 0 \\ 0 & D_-\end{array}\right)R(-\chi)
$$ 
is symmetric and describes the area distance given by $\sqrt{\det\Sg}=\sqrt{D_+D_-}$ and the shear which is parametrized by $D_+-D_-$ and $\chi$. The latter rotates the coordinate axis into the principle axes of the shear tensor. The angle $\om$ describes a rotation of the image.

In an unperturbed Friedmann Universe we have $\DD =D_A\id$ where $D_A$ denotes the  background angular diameter distance. 
 At the perturbative level, we can therefore parametrize it as
\be
\DD = D_A\left[\id +\Acal \right] \, \quad \mbox{with }~~   \Acal = \left(\begin{array}{cc} -\ka-\ga_1 ~&~ -\ga_2-\om \\ -\ga_2+\om ~&~ -\ka+\ga_1\end{array}\right)\,.
\ee 
At first order $\ka$ describes the convergence of light rays, $\ga\equiv \ga_1+i\ga_2$ describes their shear and $\om$ their rotation. 
The matrix $\Acal$ is also called the amplification matrix. The determinant $\det \DD^{-1}$ is proportional to the luminosity of the image so that, to first order in the perturbations $1+2\ka$ is the magnification~\cite{Schneider:1992}.

In GLC gauge the Jacobi map is expressed in the Sachs basis \cite{Fanizza:2013doa} so that there is no intrinsic distinction between $\beta$ and $\omega$ and one calculates directly the physical angle $\al$ which describes the rotation of an image with respect to the Sachs Basis.

Even though algebraically somewhat more involved, GLC gauge is geometrically more intuitive. Nevertheless, in the following we shall perform all the calculations in longitudinal gauge which is more commonly known.

\subsection{Scalar perturbations}\label{ss:S}

As the rotation angle $\alpha$ is an observable (hence gauge invariant) we can perform our calculations in any gauge. In longitudinal gauge,
\be
ds^2 =-a^2(\eta)(1+2\Psi)d\eta^2 + a^2(\eta)(1-2\Phi)\de_{ij}dx^idx^j \,,
\ee
where $\Phi$ and $\Psi$ are the so-called Bardeen potentials, $\eta$ denotes conformal time and, for simplicity, we have set spatial curvature to zero.
The Jacobi map for scalar perturbations at first order can be expressed in terms of the lensing potential $\psi$ given by
\be
\psi(\bx,z) = \frac{1}{2}\int_0^{r(z)}dr \frac{r(z)-r}{r(z)r}(\Phi+\Psi) \,,
\ee
where the Bardeen potentials are to be evaluated along the (unperturbed) photon geodesic and $r(z)$ is the comoving distance to redshift $z$.
The Jacobi map at first order  in these coordinates is very well known, but for completeness we here  repeat the result found in the literature, see e.g.~\cite{RuthBook}
\bea
\ka &=& \lap\psi \,,\\
\ga &=& [(\nabla_1\nabla_1 - \nabla_2\nabla_2) +2i\nabla_1\nabla_2]\psi \,,\\
\om &=& 0 \,.
\eea

To determine the rotation $\beta$ of the Sachs basis we have to integrate the geodesic transport equation.
A short calculation gives  that for a vector normal to the photon direction we have
\bea
\frac{de_a^i}{d\la}  &=& e_a^i\frac{d\Phi}{d\la}  + k^i(\bnabla\Phi\cdot{\bf e}_a)\,.
\eea
The first term just ensures that $\bf e$ remains normalized and the second term ensures the constancy of the scalar product of $\bf k$ and $\bf e$. But the basis vector ${\bf e}_a$ does not acquire any component in direction ${\bf e}_b$. Hence there is no rotation of the Sachs basis in longitudinal coordinates, $\beta=0$. This result  remains true for a quasi-Newtonian gravitational potential, $\Psi=\Phi$ non-perturbatively when replacing $1+2\Psi \mapsto \exp(2\Phi)$ and $1-2\Phi\mapsto \exp(-2\Phi)$. In longitudinal gauge the Sachs basis is not rotated with respect to the coordinate basis.

More precisely (see also~\cite{Mthesis} for a detailed discussion of this point), neglecting the time dependence of $\Phi$ and ignoring the scale factor which does not affect the conformally invariant photon geodesics, we set
\be
ds^2 = -e^{2\Phi(\bx)} d\eta^2 + e^{-2\Phi(\bx)}\de_{ij} dx^idx^j \,.
\ee
The (exact) non-vanishing Christoffel symbols of this metric are
\bea \label{eq3.38}
\Ga^0_{0i} &=&\dd_i\Phi\,, \quad \Ga^ i_{00}=\dd_i\Phi\,, \nonumber \\ 
\Ga^ i_{jm} &=& \de_{jm}\dd_i\Phi -\de^i_j\dd_m\Phi -\de^i_m\dd_j\Phi \,. 
\eea
Denoting the photon 4-vector by $k=\nu(-1,\bn)$ 
and ${e}=(e^0,\bfe)$ with $\bfe\cdot\bn=0$, parallel transport, $\nabla_ke=0$ implies
\bea\label{e:para}
\frac{de^i}{d\la} &=& e^i(\bnabla\Phi\cdot\bn) + n^i(\bnabla\Phi\cdot\bfe) \,,\\
\frac{de^0}{d\la} &=& (\bnabla\Phi\cdot\bfe)\,.
\eea
The first term ensures that the length of $e$ remains constant and the second term together with the second equation ensure that the scalar product $k^\mu e_\mu$ remains constant, and we have already made use of $\de_{ij}n^ie^j=0$. 
But clearly, $\bf e$ does not rotate in the plane normal to $u\propto \dd_\eta$ and $\bn$. Actually, the polarisation is not parallel transported but we have to project  \eqref{e:para} into the plane normal to $ u$ and $ k$ which simply removes the component in direction $\bn$ and the component $e^ 0$. Therefore the true evolution equation for the polarisation in longitudinal gauge is
\be\label{e:pol-evo}
\frac{de^i}{d\la} = e^i(\bnabla\Phi\cdot\bn)  =e^ i\frac{d\Phi}{d\la} \,. 
\ee

To summarize, $\alpha=\omega+\beta$ is equal to zero for scalar perturbation to first order.  This agrees with the result obtained in \cite{Marozzi:2016qxl} where the GLC gauge is used and the result $\al_{GLC}^{(1)}=0$ is obtained directly. 


\subsection{Vector and Tensor perturbations}\label{ss:T}

We now consider linear vector and tensor perturbations. As photon geodesics are conformally invariant, we can ignore the scale factor of the expanding universe in this calculation and consider a perturbed Minkowski metric.  
Vector and tensor perturbations in the metric then are given by
\be\label{e:VTmetric}
ds^2 =-d\eta^2 -2 B_idx^i d\eta+ (\de_{ij}+2h_{ij})dx^idx^j \,.
\ee
where $h_{ij}=\pa_{(i}F_{j)}+H_{ij}$, with $B_i$ and $F_i$ pure transverse vector perturbations and $H_{ij}$ are the  symmetric, traceless and transverse tensor perturbations and the parentheses in $\pa_{(i}F_{j)}$ denote symmetrization. The condition for the geodesic transport of the polarization $\epsilon$ can be written in full generality as (assuming $\epsilon^0 =0$)
\be
\frac{d\epsilon^i}{d\lambda}=-k^\mu{\Gamma_{\mu j}}^i \epsilon^j\equiv-K^i_j\epsilon^j
\label{eq:general_Sachs_tr}
\ee
At linear order for the metric (\ref{e:VTmetric}) we obtain
\bea
\frac{d\epsilon^i}{d\lambda}&=&-\left[ \frac{dh^i_j}{d\lambda}+k^m\left( \pa_j h_m^i-\pa^i h_{jm} \right) \qquad
\nonumber \right. \\
&& \left. \qquad \qquad
+\frac{k^0}{2}\left( \pa^i B_j-\pa_j B^i \right) \right]\bar\epsilon^j\nonumber\\
&=&-\delta^{il}\left[ \frac{dh_{lj}}{d\lambda}+k^m\left( \nabla_j h_{ml}-\nabla_l h_{jm} \right) \right. \qquad
\nonumber \\ 
&& \left.
+\frac{k^0}{2}\left( \nabla_l B_j-\nabla_j B_l \right) \right]\bar\epsilon^j
\equiv-K^i_j\bar\epsilon^j
\label{eq:356}
\eea
where $\bar\epsilon^j$ is the background direction of the polarization and $K$ now denotes the linearized expression. Notice that in the last equal sign above we changed the ordinary derivatives with the covariant ones which does not affect the result due to the antisymmetric structure of the involved terms. Without perturbations, the polarization will not rotate. The rotation we are interested in can be evaluated (always at linear order, i.e.~for small angles) as
\be
\beta=\epsilon_{ijm}\bar\epsilon^i\,\epsilon^j\,k^m
\ee
where the affine parameter $\lambda$ is normalized such that  $k^\mu=(-1,n^i)$ 
is the background direction of propagation. Then the evolution equation for $\beta$ is
\be
\frac{d\beta}{d\lambda}=\epsilon_{ijm}\bar\epsilon^i\frac{d\epsilon^j}{d\lambda}k^m
=-\epsilon_{ijm}\bar\epsilon^iK^j_l\bar\epsilon^lk^m \,.
\ee
The first term of Eq. \eqref{eq:356}, $(dh_{ij}/d\la)\bar e^j$, just integrates to $h_{ij}^{\rm{fin}}\bar e^j -h_{ij}^{\rm{in}}\bar e^j$. While this may induce a rotation it is very small, much smaller than term involving spatial derivatives and we neglected that in our treatment\footnote{This is in line with the results found in \cite{Faraoni:2007uh,Thomas:2016xhb}.}. Here we only consider the terms with the highest number of transversal derivatives since only these can contribute appreciably. With this additional approximation $K$ becomes anti-symmetric and we have
\bea
\hspace*{-0.9cm}K^j_m
&=k^l\!\left(\! \nabla_m h_l^j-\nabla^j h_{ml} \!\right)
+\frac{k^0}{2}\!\left(\! \nabla^j B_m-\nabla_m B^j \!\right) .
\label{eq:vecbeta}
\eea
It is more convenient to write the result in polar coordinates, where $k^i=\delta^i_r$ and $\bar\epsilon^i=\delta^i_a$ (latin indices $a,b,c,d$ denoting angular directions) such that we have
\beq
\frac{d\beta}{d\lambda}=-\epsilon_{abr}\bar\epsilon^aK^b_c\bar\epsilon^c \,.
\eeq
Because $K_{bc}$ is  antisymmetric we can write $K_{bc}=\epsilon_{bc}\epsilon_a^d K_d^a/2$. This leads to
\be
\bar\epsilon^i k^l\epsilon_{il}^jK_{jm} \bar\epsilon^m
=\bar\epsilon^a\epsilon_a^b\,\epsilon_{bc}\bar\epsilon^c\epsilon_e^d K_d^e/2=\epsilon_a^d K_d^a/2 \,,
\ee
for these  equalities we use that both $\bar\epsilon$  and $\bar k^i$ are normalized to 1 and  $\bar\epsilon$ is orthogonal to $\bar k^i = \delta^{ir}$. 
Hence, the rotation angle of the polarization is simply given by
\bea\label{e:beta}
\frac{d\beta}{d\lambda}
&=&-\left(\epsilon_a^d\nabla_d h_r^a
+\frac{1}{2}\epsilon_a^d\nabla_d B^a \right) \,,
\eea
where, in polar coordinates
\bea
h_{rr}&=&\nabla_r F_r+H_{rr}\,,\nonumber\\
h_{ra}&=&\nabla_{(a}F_{r)}+H_{ra} \,.
\eea

On the other hand, the image is Lie transported and the related rotation can be evaluated as the leading part in the number of spatial derivatives of the antisymmetric part of amplification matrix. This is given by
\be
\omega=\frac{1}{2}\epsilon^c_a\mathcal{A}^a_c=\frac{1}{2}\epsilon^c_a\nabla_c\theta^a \,,
\ee
where \cite{BenDayan:2012wi,Dai:2013nda,Fanizza:2018qux} 
\bea
\theta^a&=&\int_0^{\lambda_s}d\lambda\left[ B^a+2h^{ra}+\bar\ga^{ab}\pa_b\int_0^\lambda d\lambda'\left( h^{rr}+B^r \right) \right]\nonumber\\
&=&\int_0^{\lambda_s}d\lambda\left[ B^a+2h^{ra}+\bar\ga^{ab}\nabla_b\int_0^\lambda d\lambda'\left( h^{rr}+B^r \right) \right] \nonumber \\
\eea
with $a,b$ denote the angular coordinates and $r$ is the radial index and $(\bar\ga^{ab})=r^{-2}\text{diag}\left( 1,\sin^{-2}\theta \right)$. The double integral gives a symmetric contribution to $\nabla_c\theta_a$ and does therefore not contribute to the rotation so that $\omega$ is given by
\be
\omega=\int_0^{\lambda_s}d\lambda\left[ \frac{1}{2}\epsilon^c_a\nabla_cB^a+\epsilon^c_a\nabla_ch_r^a \right]
\ee
or, equivalently
\be
\frac{d\omega}{d\lambda}=\left(\frac{1}{2}\epsilon^c_a\nabla_cB^a+\epsilon^c_a\nabla_ch_r^a \right)
\ee
which agrees with the result for  $-\beta$  given in Eq.~(\ref{e:beta}). With the initial condition $\beta(0)=\om(0)=0$, this implies that $\alpha =\beta+\om =0$ for linear vector and tensor perturbations.

As for the scalar case, this result can also be obtained using  GLC gauge. As mentioned above, in this gauge we directly evaluate $\al^{(1)}$ which can easily been shown to vanish also for vector and tensor perturbations, see  Appendix C of \cite{Marozzi:2016qxl}, where this is shown in general, without decomposition into scalar vector and tensor perturbations.

This result disagrees with the  analysis presented in Ref.~\cite{Dai:2013nda}, while it is in line with Refs.~\cite{Fleury:2015hgz,Yoo:2018qba} and, regarding tensor perturbations, with Ref.~\cite{Faraoni:2007uh}. Indeed, in Ref.~\cite{Dai:2013nda} the Author expresses the polarisation rotation $\beta$ with respect to some global coordinate basis. Nevertheless this arbitrary coordinate basis is not Lie transported from the last scattering surface to the observer and, therefore, $\beta$ alone does not represent a physical, measurable rotation angle. In the analysis of  Ref.~\cite{Dai:2013nda}, the contribution of {$\omega$} has not been taken into account.


\section{Lie transport and parallel transport at second order}
\label{Sec4}

 The value for $\alpha$ was already computed in \cite{Marozzi:2016qxl} for scalar perturbation up to second order.
Here we show the computation in longitudinal gauge for convenience of the reader and also to demonstrate the gauge invariance of the result. 

As we have seen in the previous section, in  longitudinal gauge parallel transport does not lead to any rotation. However, the geodesic deviation equation which is equivalent to Lie transport does induce a non-vanishing $\om$ { in longitudinal gauge}.

 The evaluation of $\omega$ to second order for scalar perturbations has already been presented in the literature. For example, considering Eqs. (C.35)-(C.40) of~\cite{Marozzi:2016qxl} we find the following expression for $\omega$ 
\begin{widetext}
\beq \label{eq_beta2}
\om^{(2)} \left( \bx \right) =
 \frac{2}{\left( 2 \pi \right)^2} \int_0^{r_s} dr \frac{r_s - r}{r_s r} \int_0^r dr_1 \frac{r-r_1}{r r_1} 
 \int d^2 \ell_1 d^2 \ell_2 \bn \cdot \left( \Bell_2 \wedge \Bell_1 \right) \left( \Bell_1 \cdot \Bell_2 \right) \Phi_W \left( z , \Bell_1 \right) \Phi_W \left( z_1 , \Bell_2 \right) e^{-i \left( \Bell_1 + \Bell_2 \right) \cdot \bx} \, .
\eeq
Here $ \Phi_W=(\Phi+\Psi)/2$ is the Weyl potential, and $z$ and $z_1$ denote the redshift out to comoving distance $r$ and $r_1$ respectively. The comoving distance to the last scattering surface is denoted $r_s$.
Fourier transforming Eq.~\eqref{eq_beta2} we find
\be
\om^{(2)} \left( \Bell \right) = \frac{2}{\left( 2 \pi \right)} \int_0^{r_s} dr \frac{r_s - r}{r_s r} \int_0^r dr_1 \frac{r-r_1}{r r_1} 
\int d^2 \ell_1  \bn \cdot \left( \Bell \wedge \Bell_1 \right) \left( \Bell_1 \cdot \Bell - \ell_1^2 \right) \Phi_W \left( z(r) , \Bell_1 \right) \Phi_W \left( z(r_1) , \Bell -\Bell_1\right)  \, .
\ee
From this we can compute the power spectrum of the rotation angle at second order
\bea
\langle \om^{(2)} \left( \Bell \right)  \om^{(2)} \left(  \Bell' \right) \rangle &=&  \frac{4}{\left( 2 \pi\right)^2}
\int_0^{r_s} dr \frac{r_s - r}{r_s r} \int_0^r dr_1 \frac{r-r_1}{r r_1} 
\int_0^{r_s} dr \frac{r_s - r'}{r_s r'} \int_0^{r'} d{r'}_1 \frac{r'-{r'}_1}{r' {r'}_1} 
\nonumber \\
&&
\int d^2 \ell_1 d^2 \ell_2  \bn \cdot \left( \Bell \wedge \Bell_1 \right) \left( \Bell_1 \cdot \Bell - \ell_1^2 \right) \bn \cdot \left( \left( - \Bell \right) \wedge \Bell_2 \right) \left( - \Bell_2 \cdot \Bell - \ell_2^2 \right)
\nonumber \\
&&
\left[ 
C^W_{\ell_1} \left(z,z' \right)  C^W_{| \Bell - \Bell_1 | } \left( z_1 ,z'_1 \right) \delta_D \left( \Bell_1+ \Bell_2 \right) \delta_D\left(  \Bell  + \Bell' \right) \right. 
\nonumber \\
&&
\left.
+ C^W_{\ell_1} \left(z,z'_1\right)  C^W_{| \Bell - \Bell_1 | } \left( z_1 ,z' \right)
 \delta_D \left( \Bell_1- \Bell -\Bell_2 \right) \delta_D \left(  \Bell + \Bell'  \right) 
\right]
\nonumber \\
&=& \delta_D \left( \Bell +\Bell' \right)
\frac{4}{\left( 2 \pi\right)^2}
\int_0^{r_s} dr \frac{r_s - r}{r_s r} \int_0^r dr_1 \frac{r-r_1}{r r_1} 
\int_0^{r_s} dr \frac{r_s - r'}{r_s r'} \int_0^{r'} d{r'}_1 \frac{r'-{r'}_1}{r' {r'}_1} 
\nonumber \\
&&
\int d^2 \ell_1 \left[\bn \cdot \left( \Bell \wedge \Bell_1\right) \left( \Bell_1 \cdot \Bell - \ell_1^2 \right)\right]^2
\left[ C^W_{\ell_1} \left(z,z' \right)  C^W_{| \Bell - \Bell_1 | } \left( z_1 ,z'_1 \right) -  C^W_{\ell_1} \left(z,z'_1\right)  C^W_{| \Bell - \Bell_1 | } \left( z_1 ,z' \right) \right]\, . \qquad
\eea
Inserting $\langle {\om}^{(2)} \left( \Bell \right) {\om}^{(2)} \left(  \Bell' \right) \rangle= \delta_D \left( \Bell +\Bell' \right) C^{\om \om}_\ell $  and denoting the transfer function of the Weyl potential $T_{\Phi + \Psi}(k,z)$, we obtain \cite{Marozzi:2016qxl}, with the help of the Limber approximation~\cite{Limber:1954zz,LoVerde:2008re}, the result
\bea \label{clbetabeta}
 C^{\om \om}_\ell &=&
\frac{1}{4 \left( 2 \pi \right)^2} \int_0^{r_s} \frac{dr}{r^2} \int_0^r \frac{dr_1}{r_1^2} \left( \frac{r-r_1}{r r_1} \right)^2\left( \frac{r_s - r}{r_s r} \right)^2
\int d^2\ell_1  \left[\bn \cdot \left( \Bell \wedge \Bell_1\right) \left( \Bell_1 \cdot \Bell - \ell_1^2 \right)\right]^2
\nonumber \\
&&
\left[ T_{\Phi + \Psi} \left( \frac{\ell_1 +1/2}{r} , z \right) T_{\Phi + \Psi} \left( \frac{| \Bell - \Bell_1 | +1/2}{r_1} , z_1 \right) \right]^2 
P_R \left( \frac{\ell_1 + 1/2}{r} \right) P_R \left( \frac{| \Bell - \Bell_1 |+1/2 }{r_1} \right)
=\, C^{\al \al}_\ell\, ,
\eea
 where $P_R\left( k \right)$ is the primordial curvature power spectrum.
For the last equal sign we used that $\beta=0$ in longitudinal gauge.
\end{widetext}

 \begin{figure}[t]
\centering
\includegraphics[width=1\linewidth]{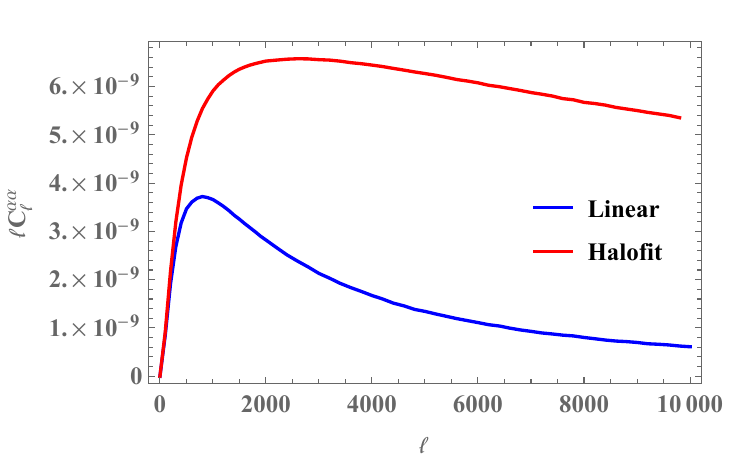}
\\
\includegraphics[width=1\linewidth]{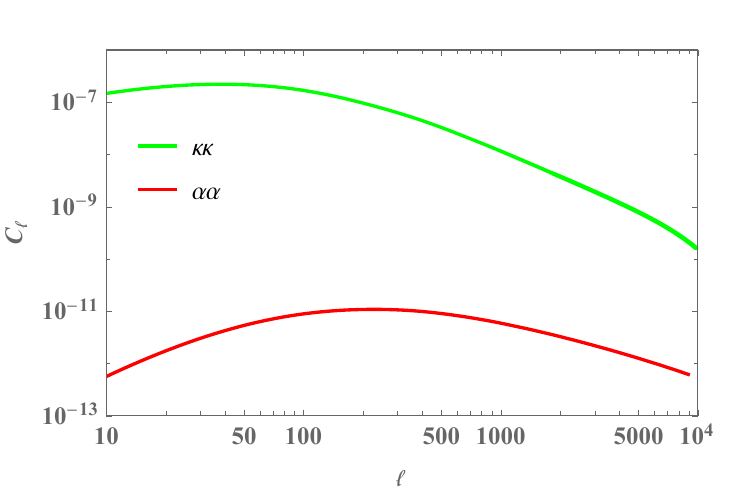}
\caption{Top panel: we plot the angular power spectrum of the rotation angle $\alpha=\om_{(LG)}$. In blue by using the linear power spectrum and in red with Halofit. Bottom panel: as comparison we show the angular spectrum of the rotation angle (red) together with the spectrum of the convergence $\kappa$ (green). The first is related to the curl potential as $C_\ell^{\alpha \alpha} = \ell^4 C_\ell^{\Omega\Omega} /4$, while the latter to the lensing potential $\phi$ through $C_\ell^{\kappa\kappa} = \ell^4 C_\ell^{\phi \phi} /4$.}
\label{fig:omspec}
\end{figure}

The numerical results for $\tilde{C}_{\ell}^{(Br)}$ have been generated by performing the double integral \eqref{rot_B_eq} with $F_{\ell \ell' \ell''}$ given in \eqref{F_factor2} using the same cosmological parameters as Ref.~\cite{Marozzi:2016qxl} for comparison purpose. Namely  $h = 0.67$, $h^2\Omega_{\rm cdm}=\omega_{\rm cdm} = 0.12$, $\Omega_bh^2=\omega_b = 0.022$ and vanishing curvature. The primordial curvature power spectrum has the amplitude $A_s = 2.215 \times 10^{-9}$ at the pivot scale $k_{\rm pivot} = 0.05 {\rm Mpc}^{-1}$, the spectral index $n_s =0.96$ and no running is assumed. The transfer function for the Bardeen potentials, $T_{\Phi +\Psi}$ has been computed with \class{}~\cite{Blas:2011rf} using the linear power spectrum and Halofit~\cite{Takahashi:2012em}.

 \begin{figure}[t!]
\centering
\includegraphics[width=1\linewidth]{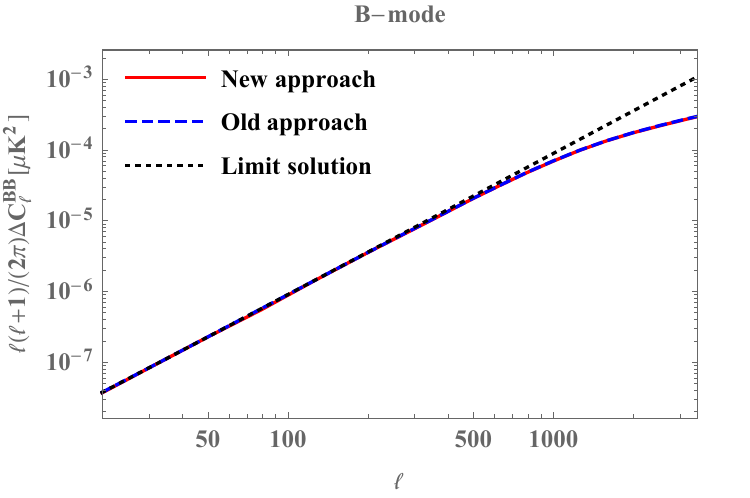}\vspace{0.5cm}
\includegraphics[width=1\linewidth]{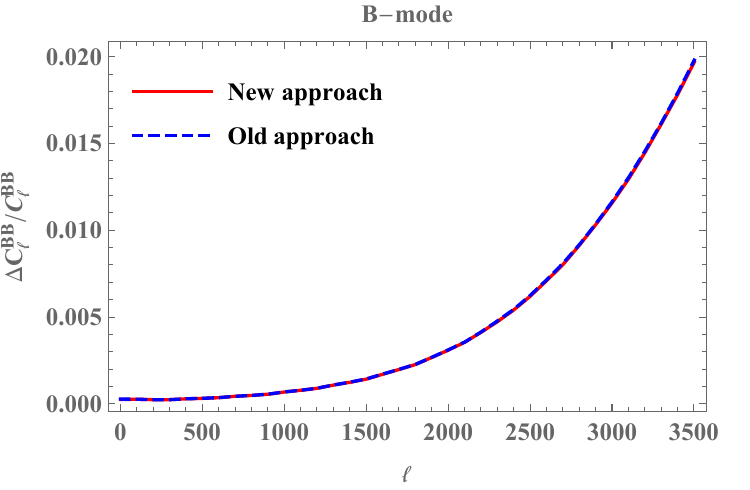}
\caption{We show the effect induced by rotation on the angular $B$-mode power spectrum. In the top panel we show the good accuracy of the low-$\ell$ limit solution derived in Eq. \eqref{limit_sol}. In the bottom panel we show the relative amplitude compared to the first order lensed $B$-mode. { The red line is the present result according to Eq.~\eqref{F_factor2}, the dashed blue line refers to our previous result~\cite{Marozzi:2016qxl} recomputed by integrating Eq.~\eqref{eq.b2} and the dotted black line is the limit solution described by Eq.~\eqref{limit_sol}.}} \label{f:Brotspec}
\end{figure}

From Fig.~\ref{fig:omspec} we see that the lensing spectrum increases by about a factor 5 on small scales when using the non-linear Halofit spectrum and $\ell C_\ell^{{\al}}$ decays very slowly with $\ell$. In Appendix~\ref{app:B} we also show the formal equivalence of the expression \eqref{rot_B_eq} and the result obtained in \cite{Marozzi:2016qxl}.
In Fig.~\ref{f:Brotspec} we plot the $B$-mode power spectrum induced from rotation of polarisation (top panel). In the lower panel we plot the relative contribution to the first order lensing $B$-spectrum. As a numerical cross-check we show also the results by integrating the double integral given by Eq.~\eqref{eq.b2} and the low $\ell$ approximation given in  Eq.~\eqref{limit_sol}.

 \begin{figure}[ht!]
\centering
\includegraphics[width=1\linewidth]{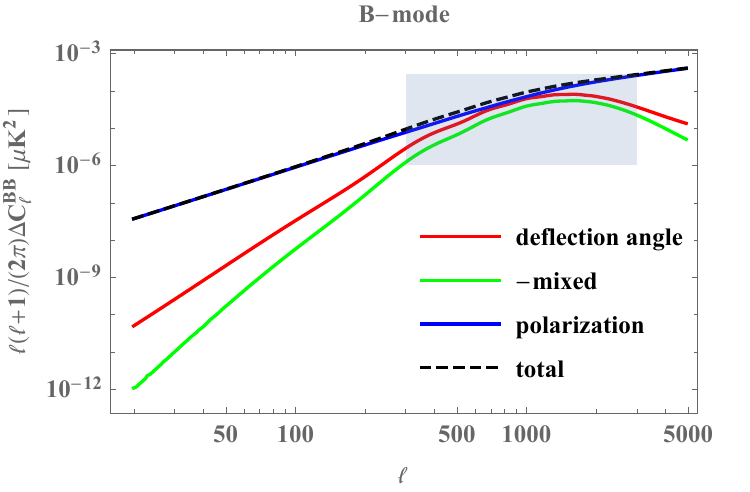}
\includegraphics[width=1\linewidth]{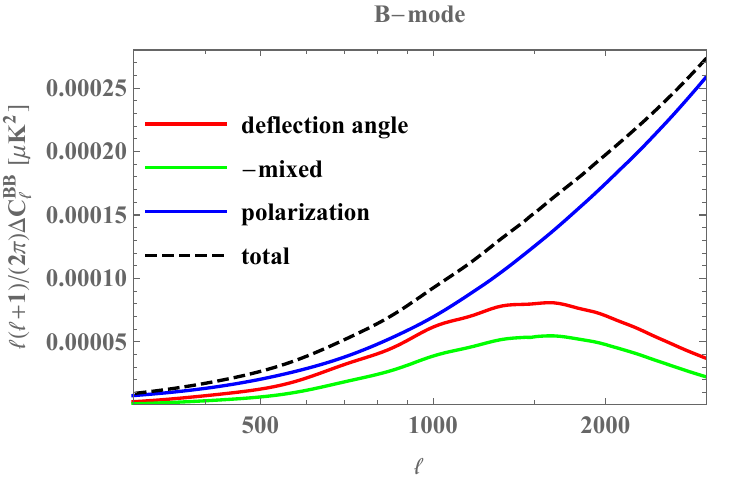}
\caption{We show all three  contributions to $\De C_\ell^B$, the polarisation rotation (blue line), which is also shown in Fig.~\ref{f:Brotspec}, the curl deflection (red line) and the negative of the mixed term (green line). Their sum is indicated as dashed black line. The bottom plot is a magnification of the gray region of the top panel, in order to compare the different effects at the scales where they are comparable.\label{f:Ball}}
\end{figure}

In Fig.~\ref{f:Ball} we show the different contributions including also the curl-type deflection angle term computed in \eqref{e:Bcurl} and the mixed term~\eqref{e:mixed}. Cleary the two additional terms are relevant mainly around $\ell\sim 1000$, where they  amount to about $25\%$ of the total result. 

In Appendix~\ref{app:C} we explain the shape of the three terms in detail.

\section{Discussion and conclusion}\label{Sec5}

In this paper we clarify an issue concerning the rotation of polarisation under the parallel transport of CMB photons in the clustered Universe. We show that the relevant angle is the one between parallel transported vectors and geodesic deviation vectors which are Lie transported. Or, in other words, the rotation of the geodesic deviation vector in the Sachs basis. This well defined geometric angle which we call $\al$  vanishes at first order, but not at second order. Its second order value is therefore gauge-invariant as a consequence of the Stewart-Walker lemma \cite{Stewart:1974uz} and its generalization to higher-order \cite{Bruni:1996im}. Denoting the angle of rotation of the Sachs basis (with respect to some arbitrary coordinate basis) by $\beta$ and the one of geodesic deviation vectors  (with respect to the same arbitrary basis) by $\om$ we have $\al=\beta+\om$. 
 For scalar perturbation, we have shown that in longitudinal gauge, $\beta=0$ at all orders. The gauge invariance of $\al^{(2)}$ is confirmed by the finding that $\om^{(2)}_{LG}=\al^{(2)}_{GLC}$.

Even if observers measure polarisation with respect to a fixed observer coordinate system (they measure the Stokes parameters), they then combine the coordinate dependent Stokes parameters into the coordinate independent $E$- and $B$-polarisation spectra and these are affected by rotation in the way computed here.

This result is important for polarisation measurements with high sensitivity, like CMB S4~\cite{Abazajian:2016yjj}, which want to detect primordial gravitational waves with a tensor-to-scalar ratio as small as $r\sim 10^{-3}$. To correctly subtract the lensing contribution to the $B$-polarisation this requires a precision of better than 0.1\% for the lensing spectrum in the crucial $\ell$ range which is used for de-lensing, namely $1000\leq \ell \leq 3000$. But in this $\ell$ range the contribution from rotation increases
up to 1\% and therefore has to be considered. 

The amplitude of the effects induced by the curl component $\Omega$ (with $\alpha= - \Delta \Omega/2$) could reduce the efficiency of de-lensing gradient based methods~\cite{Hirata:2003ka}. This may set an accuracy limit in the search for primordial B-modes and, in general, weaken the constraints on cosmological parameters strongly sensitive to the sharpness of BAO peaks in the CMB power spectrum (that are smeared out by lensing), like e.g. neutrino masses.

Furthermore, even if $r$ is much smaller than what an experiment can ever reach, measuring the rotation of polarisation is a measurement of frame dragging on cosmological scales which would represent a formidable test of General Relativity on these scales.

\section*{Acknowledgements}
We thank Camille Bonvin, Anthony Challinor, Chris Clarkson, Pierre Fleury,  Alex Hall, Martin Kunz, Antony Lewis, Roy Maartens, Miguel Vanvlasselaer and Gabriele Veneziano for helpful and clarifying discussions.
RD is grateful for the hospitality and for financial support of the Physics Department and INFN section of Pisa. GF is grateful for the hospitality of the Physics Department of the University of Bari. ED (No.~171494) and RD acknowledge support from the Swiss National Science Foundation.
GF and GM are supported in part by INFN under the program TAsP (Theoretical Astroparticle Physics).

\newpage
\appendix
\begin{widetext}
\section{Details on the geometrical factor $F_{k \ell q}$} \label{app:A}
In order to evaluate analytically the geometrical factor $F_{}$ defined in Eq.~\eqref{F_factor}, we use the following identity~\cite{Asgari:2016txw}, see also~\cite{GradRi},
\be \label{3bessel_int}
\int_0^\infty ds s J_0 \left( q s \right) J_n \left( k s \right) J_n \left( \ell s \right) = \text{Re}\left(\frac{\cos \left( n \theta \right) }{\pi k \ell \sin \theta}\right)
\,,
\ee
where 
\be
\cos \theta = \frac{\ell^2 + k^2 - q^2}{2 k \ell}\, .
\ee
The real part 'Re' ensures that the integral vanishes if $(q,k,\ell)$ do not satisfy the triangle inequality, and $\theta$ is the angle between the sides of lengths $k$ and $\ell$ in the triangle formed by $(q,k,\ell)$.
In particular, we are interested in the following integrals
\bea
 \int_0^\infty ds s J_0 \left( q s \right) J_0 \left( k s \right) J_0 \left( \ell s \right) &=&
\frac{2 \text{Re}}{\pi } \left(\frac{1}{\sqrt{(q^2-(k-\ell)^2) ((k+\ell)^2-q^2)}}\right) \, ,
\\
 \int_0^\infty ds s J_0 \left( q s \right) J_4 \left( k s \right) J_4 \left( \ell s \right) &=&
 \frac{\left(k^8-4 k^6 q^2+k^4 \left(6 q^4-4 \ell^2 q^2\right)-4 k^2 q^2 \left(\ell^2-q^2\right)^2+\left(\ell^2-q^2\right)^4\right)}{\pi  k^4 \ell^4}
 \nonumber \\
 && \times
  \text{Re}\left(\frac{1}{\sqrt{(q^2-(k-\ell)^2) ((k+\ell)^2-q^2)}}\right) \, .
\eea
At the boundaries of the triangle equality, i.e. $q=|k\pm\ell |$, the integrals diverge and they need to be interpreted as a distribution within integral~\eqref{rot_B_eq}.  With this identity we can rewrite the geometrical factor~\eqref{F_factor} as
\be \label{F_factor2}
F_{ k \ell q} = \frac{\left(k^4-2 q^2 \left(k^2+\ell^2\right)+\ell^4+q^4\right)^2 \text{Re}\left(\frac{1}{\sqrt{(q^2-(k-\ell)^2) ((k+\ell)^2-q^2)}}\right)}{\pi  k^4 \ell^4}\, .
\ee

\section{Equivalence with the previous calculation} \label{app:B}

In this appendix we show that the expression for the $B$-mode induced by rotation from our previous paper (see Eq.~(6.17) in Ref.~\cite{Marozzi:2016qxl}) is equal to Eq.~\eqref{rot_B_eq}.  We start with 
Eq.~(6.17) of Ref.~\cite{Marozzi:2016qxl}
\bea
\Delta C^{B (2,2)}_\ell&\equiv&\frac{1}{2}\left[ \Delta\left( C^\Ecal_\ell+C^\Bcal_\ell \right)^{(2,2)}-\Delta\left( C^E_\ell-C^B_\ell \right)^{(2,2)}\right]\nonumber\\
&=&16\,\int \frac{d^2\ell_1}{(2\pi)^2}\,\int \frac{d^2\ell_2}{(2\pi)^2}
\left[\boldsymbol{n}\cdot\left( \boldsymbol{\ell}_2\land\boldsymbol{\ell}_1 \right)\left(\boldsymbol{\ell}_1\cdot\boldsymbol{\ell}_2\right)\right]^2\,\int_{0}^{r_s}dr\,\frac{r_s-r}{r_s\,r}\int_0^{r}dr_1\,\frac{r-r_1}{r\,r_1}
\nonumber\\
&&\times
\int_{0}^{r_s}dr_2\,\frac{r_s-r_2}{r_s\,r_2}\int_{0}^{r_2}dr_3\,\frac{r_2-r_3}{r_2\,r_3}
\left[ C^W_{\ell_1}(z,z_2)C^W_{\ell_2}(z_1,z_3)
-C^W_{\ell_1}(z,z_3)C^W_{\ell_2}(z_1,z_2)\right]\nonumber\\
&&\times \left\{ C^E_{|\Bell-\Bell_1-\Bell_2|}(z_s)\cos^2\left[2 \left( \varphi_\ell - \varphi_{|\Bell-\Bell_1-\Bell_2|} \right)\right]\right.
\left.+C^B_{|\Bell-\Bell_1-\Bell_2|}(z_s)\sin^2\left[2 \left( \varphi_\ell - \varphi_{|\Bell-\Bell_1-\Bell_2|} \right)\right] \right\}\,,
\label{B22}
\eea
then, by making a change of variable $\Bell_2= \Bell' - \Bell_1 $ and using Eqs.~\eqref{clbetabeta}, we obtain (in absence of primordial $B$-mode) 
\be \label{eq.b2}
\Delta C^{B (2,2)}_\ell= 4 \int \frac{d^2 \ell'}{\left( 2 \pi \right)^2} C^{\omega\omega}_{\ell'} C^E_{|\Bell - \Bell'|} \cos^2 \left( 2 \varphi_\ell - 2 \varphi_{|\Bell - \Bell'|} \right)\, .
\ee

For low $\ell$ we can approximate the contribution induced by the rotation to the $B$-mode as follows
\be \label{limit_sol}
\Delta C^{B (2,2)}_\ell \simeq \int \frac{d \ell'}{\pi} \ell' C^{\omega\omega}_{\ell'} C^E_{\ell'} + \mathcal{O} \left( \ell^2 \right)
= \int \frac{d \ln \ell'}{\pi} \ell'^2 C^{\omega\omega}_{\ell'} C^E_{\ell'}  + \mathcal{O} \left( \ell^2 \right) \sim 5 \times 10^{-10} \mu K^2\, .
\ee
As we see from the upper panel of Fig.~\ref{f:Brotspec}, this limiting white noise contribution fully captures the power-law dependence induced by the rotation up to scale $\ell\sim 1000$.

We now show that this result is equivalent to Eq.~\eqref{rot_B_eq}. Since $C_\ell$ is independent of direction we may rotate $\Bell$ such that $\varphi_\ell = 0$. We then find
\be \label{eq.B4}
\Delta C^{B (2,2)}_\ell = 4 \int \frac{d^2 \ell''}{\left( 2 \pi \right)^2} C^{\omega\omega}_{\ell''} C^E_{|\Bell - \Bell''|} \cos^2 \left( 2 \varphi_{(\Bell - \Bell'')} \right)
= 4 \int \frac{d^2 \ell''}{\left( 2 \pi \right)^2} d \ell' \delta_D \left( \ell' - |\Bell - \Bell''| \right) C^{\omega\omega}_{\ell''} C^E_{\ell'} \cos^2 \left( 2 \varphi_{(\Bell - \Bell'')} \right) \, .
\ee
We rewrite the Dirac delta distribution as
\bea \label{eq:A13}
\delta_D \left( \ell' - |\Bell - \Bell''| \right) = \delta_D \left( \ell' - \sqrt{\ell^2 + {\ell''}^2 - 2 \ell \ell'' \cos \varphi_{\ell''}} \right) =\sum_{i=1}^2 \delta_D  \left( \varphi_{\ell''} - \varphi_i \right) \left| \frac{\sqrt{\ell^2 + {\ell''}^2 - 2 \ell \ell'' \cos \varphi_i}}{\ell \ell'' \sin \varphi_i}\right| \,,
\eea
where 
\be
\varphi_{1,2}= \pm \arccos\left(\frac{\ell^2-{\ell'}^2+{\ell''}^2}{2 \ell {\ell''}}\right) \,,
\ee
 if $ |\ell^2-{\ell'}^2+{\ell''}^2| < 2 \ell  \ell''$. If the triangle equality is not satisfied, the integral vanishes.
The triangle inequality can be explicitly enforced by writing
\bea
&&\int d\varphi_{\ell''} f\left( \varphi_{\ell''}\right)
\delta_D \left( \ell' - |\Bell - \Bell''| \right)  =  
\nonumber \\
&& =
2 \ell' \text{Re} 
\left[ \frac{1}{\sqrt{{\ell'}^2 - \left( \ell-\ell''\right)^2 }  \sqrt{ \left( \ell+\ell''\right)^2 -{\ell'}^2 }}
\right] 
\int d\varphi_{\ell''}f\left( \varphi_{\ell''}\right)
\left[ \delta_D \left( \varphi_{\ell''}- \varphi_1 \right) + \delta_D \left( \varphi_{\ell''}- \varphi_2 \right) \right] \, .
\eea
The angular integral over $\varphi_{\ell''}$ can now be performed analytically,
\bea &&
\int d\varphi_{\ell''} \cos^2 \left( 2 \varphi_{(\Bell - \Bell'' )}\right) \left[ \delta_D \left( \varphi_{\ell''}- \varphi_1 \right) + \delta_D \left( \varphi_{\ell''}- \varphi_2 \right) \right] 
= 
\nonumber \\
&=& 2 
\int d\varphi_{\ell''} \frac{\left(\ell^2-2 \ell {\ell''} \cos (\varphi_{\ell''})+{\ell''}^2 \cos (2 \varphi_{\ell''})\right)^2}{\left(\ell^2-2 \ell {\ell''} \cos (\varphi_{\ell''})+{\ell''}^2\right)^2} \delta_D \left( \varphi_{\ell''}- \varphi_1 \right)
= \frac{\left(\ell^4-2 {\ell''}^2 \left(\ell^2+{\ell'}^2\right)+{\ell'}^4+{\ell''}^4\right)^2}{2 \ell^4 {\ell'}^4} \,.
\eea
Inserting this in (\ref{eq.B4}) we find 
\bea
\Delta C^{B (2,2)}_\ell &=& 4 \int \frac{d^2 \ell''}{\left( 2 \pi \right)^2} C^{\omega\omega}_{\ell''} C^E_{|\Bell - \Bell''|} \cos^2 \left( 2 \varphi_{|\Bell - \Bell''|} \right)
=
\nonumber \\
&=&
\frac{1}{\pi} \int d\ell' d\ell'' \ell' \ell'' C^{\omega\omega}_{\ell''} C^E_{\ell'} 
\nonumber \\
&& \times \frac{\left(\ell^4-2 {\ell''}^2 \left(\ell^2+{\ell'}^2\right)+{\ell'}^4+{\ell''}^4\right)^2}{\pi \ell^4 {\ell'}^4} 
\text{Re} \left[ \frac{1}{\sqrt{{\ell'}^2 - \left( \ell-\ell''\right)^2 }  \sqrt{ \left( \ell+\ell''\right)^2 -{\ell'}^2 }} \right]
\nonumber \\
&=&
\frac{1}{\pi} \int d\ell' d\ell'' \ell' \ell'' C^{\omega\omega}_{\ell''} C^E_{\ell'}  F_{\ell \ell' \ell''} \,,
\eea
where we have used the definition~\eqref{F_factor2} in the last equality.
\end{widetext}


\section{The shape of the spectra}\label{app:C}
\begin{figure}[ht!]
\centering
\includegraphics[width=1\linewidth]{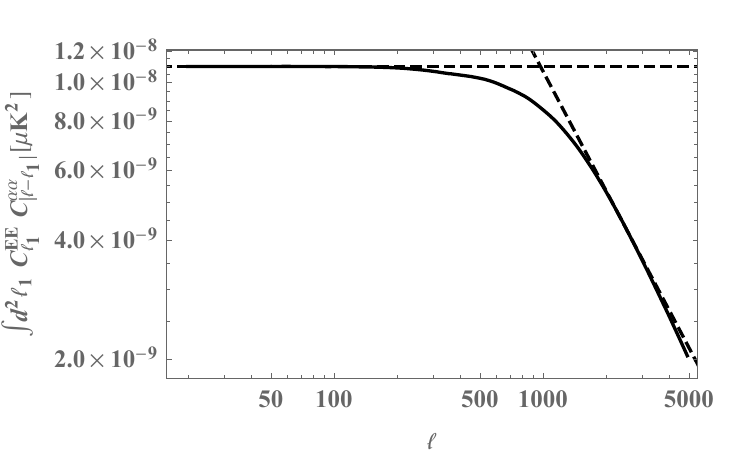}
\caption{We show the convolution spectrum $C^{\rm con}_\ell=\int d^2\ell_1 C^{\al}_{\ell-\ell_1}C^E_{\ell_1}$.  \label{f:CalCe}}
\end{figure}

To discuss the form of the spectra shown in Fig.~\ref{f:Ball}, we plot the full convolution $C^{\rm con}_\ell=\int d^2\ell_1 C^{\al}_{|\Bell-\Bell_1|}C^E_{\ell_1}$ in Fig.~\ref{f:CalCe}. This spectrum starts off as white noise and decays roughly like $\ell^{-1}$ for $\ell>800\equiv \ell_{\max}$.  This comes from the same behavior of the $ C^{\al}_\ell$ spectrum for large $\ell$. Together with equation (\ref{e:Btot}) this explains the growth $\propto \ell$ of $\ell^2\tilde C_\ell^{Br}$ at high $\ell$ and the decay of $\ell^2\tilde C_\ell^{Bd}$ like $\ell^{-1}$ due to the additional pre-factor which reduces to $1/\ell^2$ at high $\ell$. For small $\ell\ll\ell_{\max}$, $\ell^2\tilde C_\ell^{Br}\propto \ell^2$ behaves like white noise, while $\ell^2\tilde C_\ell^{Bd}$ has an additional suppression factor of roughly $(\ell/\ell_{\max})^2$ due to the $\ell$-dependent pre-factor. The amplitude at $\ell\sim \ell_{\max}\sim 10^3$ is of the order of $\ell^2\times 10^{-8}/(2\pi)^2 \sim 10^{-4} \mu K^2 $ which is in the right bull park.
\\
The mixed spectrum, $\tilde C_\ell^{Brd}$ is somewhat more intricate. The mixed term acquires naively a factor $\ell/\ell'$ for low $\ell$, but the true spectrum scales as $\ell^4$ at low $\ell$. This is due to an additional cancellation coming from positive and negative contributions in the angular integral of $\sin(4\varphi_{\ell'\ell})\sin\varphi_{\ell'\ell}$ and requires a more subtle analysis: The pure integral $\int d\varphi\sin(4\varphi)\sin(\varphi) =0$, hence the mixed contribution does not vanish only due to the angular dependence of $C^\al_{|\Bell-\Bell'|}$.  Approximating $C^\al_{|\Bell-\Bell'|}$ by a polynomial in $|\Bell-\Bell'|/\ell'$ for small $\ell$, the first non-vanishing contribution in the angular integral comes from $(\ell\cos\varphi)^3$, which increases as $\ell^3$ for small $\ell$.  This leads to a behavior of $\ell^2\tilde C_\ell^{Brd}\propto \ell^6$ for small $\ell$. The peak at $\ell\sim \ell_{\max}$ is again determined by the 'peak' of $C^\al_{|\Bell-\Bell'|}C^E_{\ell'}$ at roughly this scale.  For large $\ell\gg \ell_{\max}$, the spectrum $C^\al_{|\Bell-\Bell'|}$ goes approximately as $|\Bell-\Bell'|^{-1}$. Including also the pre-factor $\ell/|\Bell-\Bell'|^2$, again the first non-vanishing term in the angular integral comes from $((\ell'/\ell)\cos\varphi)^3/\ell^2$. We therefore expect $\ell^2\tilde C_\ell^{Brd}\propto \ell^{-3}$. These behaviors are of course quite crude approximations (e.g.~the $\al$-spectrum decays slightly faster then $\ell^{-1}$ at larger $\ell$ which is also visible in Fig.~\ref{f:CalCe}) but they reflect rather well the asymptotic slopes of the curves shown in Fig.~\ref{f:Ball}.

\bibliographystyle{unsrtnat}
\bibliography{biblio_CMBlensing}

\end{document}